\documentclass[a4paper]{spie_margins}

\usepackage{amsmath,amsfonts,amssymb}
\usepackage{graphicx}
\usepackage[colorlinks=true, allcolors=blue]{hyperref}

\usepackage{textcomp}    
\usepackage{xcolor}      
\usepackage{aas_macros}  

\graphicspath{{./images/}}  

\title{The Gravitational-wave Optical Transient Observer (GOTO)}

\author[a]{\mbox{Martin J. Dyer}}
\author[b]{\mbox{Danny Steeghs}}
\author[c]{\mbox{Duncan K. Galloway}}
\author[a,d]{\mbox{Vik S. Dhillon}}
\author[e]{\mbox{Paul O'Brien}}
\author[f]{\mbox{Gavin Ramsay}}
\author[g]{\mbox{Kanthanakorn Noysena}}
\author[d]{\mbox{Enric Pallé}}
\author[h]{\mbox{Rubina Kotak}}
\author[i]{\mbox{Rene Breton}}
\author[j]{\mbox{Laura Nuttall}}
\author[b]{\mbox{Don Pollacco}}
\author[b]{\mbox{Krzysztof Ulaczyk}}
\author[b]{\mbox{Joseph Lyman}}
\author[c]{\mbox{Kendall Ackley}}
\author[ ]{\mbox{the GOTO Collaboration}}

\affil[a]{Department of Physics and Astronomy, University of Sheffield, Sheffield S3 7RH, UK}
\affil[b]{Department of Physics, University of Warwick, Coventry CV4 7AL, UK}
\affil[c]{School of Physics \& Astronomy, Monash University, Clayton VIC 3800, Australia}
\affil[d]{Instituto de Astrofísica de Canarias, E-38205 La Laguna, Tenerife, Spain}
\affil[e]{School of Physics \& Astronomy, University of Leicester, University Road, Leicester LE1 7RH, UK}
\affil[f]{Armagh Observatory \& Planetarium, College Hill, Armagh, BT61 9DG, UK}
\affil[g]{National Astronomical Research Institute of Thailand, 260 Moo 4, T. Donkaew, A. Maerim, Chiangmai, 50180
Thailand}
\affil[h]{Department of Physics \& Astronomy, University of Turku, Vesilinnantie 5, Turku, FI-20014, Finland}
\affil[i]{Jodrell Bank Centre for Astrophysics, Department of Physics and Astronomy, The University of Manchester, Manchester M13 9PL, UK}
\affil[j]{University of Portsmouth, Portsmouth, PO1 3FX, UK}

\authorinfo{Send correspondence to MJD - email: martin.dyer@sheffield.ac.uk}

\begin{document} 
\maketitle

\begin{abstract}
The Gravitational-wave Optical Transient Observer (GOTO) is a wide-field telescope project focused on detecting optical counterparts to gravitational-wave sources. GOTO uses arrays of 40 cm unit telescopes (UTs) on a shared robotic mount, which scales to provide large fields of view in a cost-effective manner. A complete GOTO mount uses 8 unit telescopes to give an overall field of view of 40 square degrees, and can reach a depth of 20th magnitude in three minutes. The GOTO-4 prototype was inaugurated with 4 unit telescopes in 2017 on La Palma, and was upgraded to a full 8-telescope array in 2020. A second 8-UT mount will be installed on La Palma in early 2021, and another GOTO node with two more mount systems is planned for a southern site in Australia. When complete, each mount will be networked to form a robotic, dual-hemisphere observatory, which will survey the entire visible sky every few nights and enable rapid follow-up detections of transient sources.
\end{abstract}


\section{Introduction}
\label{sec:introduction}

The first detection of an electromagnetic counterpart of a gravitational-wave signal in August 2017 lead to an unprecedented global follow-up campaign, involving facilities across both the globe and the electromagnetic spectrum\cite{GW170817, GW170817_followup}. This event, GW170817, was the first signal detected from a binary neutron star merger and was localised to a $\sim$30~square~degree region on the sky, which allowed telescopes with relatively small fields of view to efficiently cover all galaxies in the search area and locate the counterpart kilonova. However, it is expected that such a small search area will be a rare occurrence, and indeed subsequent events have not been as well localised. The second binary neutron star gravitational-wave detection, GW190425, was only reliably detected by a single gravitational-wave detector, meaning the initial localisation region covered just over 10,000~square~degrees\cite{GW190425}. The challenge therefore is to search such large sky areas quickly enough to find the counterpart before it fades, while also discarding the large number of false positive detections. The need for speed and coverage points towards wide-field robotic telescopes, backed by an automated candidate-detection pipeline.

\newpage

\section{The GOTO project}
\label{sec:goto}

\begin{figure}[t]
    \begin{center}
        \includegraphics[width=0.9\linewidth]{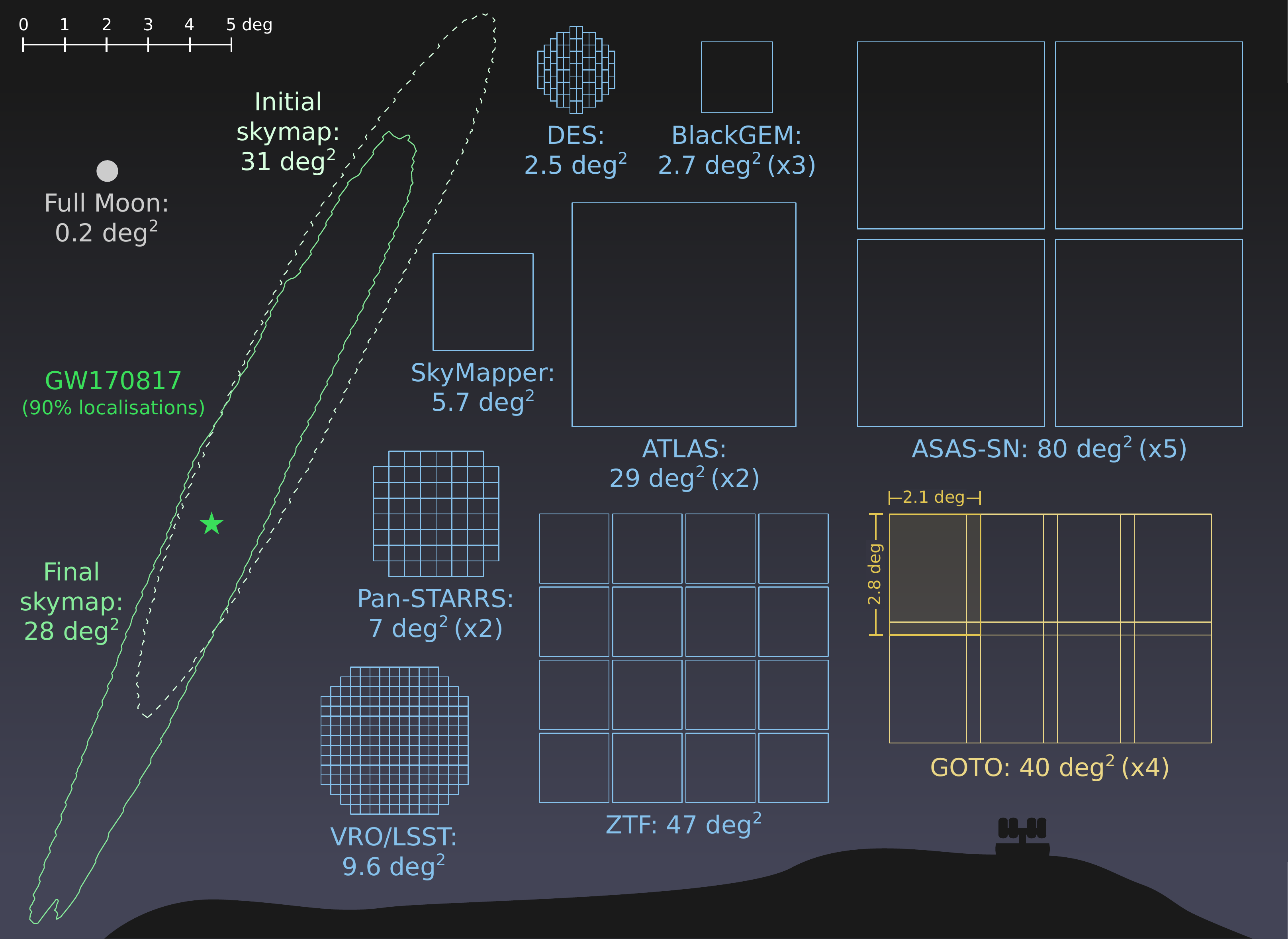}
    \end{center}
    \caption[example]{
        The field of view of a single GOTO mount (in yellow), compared to other selected wide-field telescopes (blue) and the initial and final skymaps for GW170817 (green). The complete 40~sq~deg GOTO footprint is formed from the overlapping fields of view of the 8 unit telescopes on the mount, one of which is highlighted.
    }\label{fig:fov}
\end{figure}

The Gravitational-wave Optical Transient Observer (GOTO) collaboration\footnote{\url{https://goto-observatory.org/}} was founded in 2014 and, as of 2020, contains 10 institutions from the UK, Australia, Thailand, Spain and Finland. The aim of the GOTO project is the detection of optical counterparts of gravitational-wave sources and other fast-evolving transients, through the construction of a network of robotic, wide-field telescopes.

\subsection{Design}
\label{sec:design}

There are multiple parameters to consider when designing telescopes to follow-up gravitational-wave sources. In order to maximise the chance of detecting an electromagnetic counterpart, telescopes are required to search a sufficiently large area to a sufficient depth. There is generally a trade-off between the two: traditionally telescopes with larger apertures which can observe fainter sources have small fields of view, while projects wanting to cover larger areas often use arrays of small-aperture telescopes (e.g. ASAS-SN\cite{ASAS-SN}, SuperWASP\cite{SuperWASP}). Recent projects which aim to cover both area and depth include the Zwicky Transient Facility (ZTF)\cite{ZTF} and the Vera Rubin Observatory (VRO)\cite{LSST}, but come at a significant financial cost. The third consideration is speed, as the counterpart will be rapidly fading and quick identification is required for verification by other facilities. Flexibility in scheduling and sky coverage is also required in order to efficiently cover large, irregular search regions; for this a large simultaneous field of view split across multiple mounts can be more useful than a single wide-field telescope. Finally having multiple telescopes across the globe is vital in order to cover the largest probability skymaps, in particular the ability to observe in both the northern and southern hemispheres.

In order to be able to detect the transient counterparts the telescopes have to operate in two distinct modes. When a gravitational-wave alert is received the telescopes should respond as soon as possible, ideally automatically without the need for human involvement. The regions of the sky contained within the high-probability skymap area, which could span hundreds or thousands of square degrees, then need to be covered as quickly as possible, ideally with multiple passes, which requires an efficient method of dividing the region and scheduling observations. However the images taken this way can only be used to detect transient sources if there is an existing comparison image taken previously at the same sky position. Therefore when not responding to alerts the telescopes should operate in survey mode, aiming to build up an archive of recent reference images of the whole sky. These reference templates can then be used to detect new candidate sources when an alert occurs, as well as identify false-positive detections.

\subsection{Hardware}
\label{sec:hardware}

The GOTO telescopes have been designed to minimise the response time and achieve a balance between sky coverage and depth. Each GOTO telescope is built around a fast-slewing German equatorial mount, which can hold up to 8 unit telescopes (UTs). The array design provides a cost-effective way of reaching the desired wide field of view with multiple small telescopes instead of one large one, and the modular nature also allows more unit telescopes to be added to a mount as more funding becomes available. All the unit telescopes on a mount are combined, with a small overlap, to form the overall field of view of $\sim$40~square~degrees (see Figure~\ref{fig:fov}).

The design for the GOTO unit telescopes required a large-format detector with small pixels. The On-Semi KAF-50100 CCD\footnote{\url{https://www.onsemi.com}} was chosen as it provided the best combination of small (6 micron) pixels in a large ($8176\times6132$) array for a reasonable cost. The CCDs are housed in FLI MicroLine cameras\footnote{\url{https://www.flicamera.com}}, and each UT also has a filter wheel with a set of Baader \textit{LRGBC} filters\footnote{\url{https://www.baader-planetarium.com}}. The unit telescopes themselves are fast Wynne-Newtonian astrographs, with 40cm apertures and a focal ratio of $f/2.5$. Combined with the cameras each UT therefore produces a plate scale of 1.25”/pixel and an individual field of view of $\sim$6.9~square~degrees. Three stacked 60~s images reach typical a limiting magnitudes of 19-21, depending on the sky brightness.

Each GOTO telescope is housed in an 18ft Astrohaven clamshell dome\footnote{\url{https://www.astrohaven.com}}, which when fully open allows an unrestricted view of the sky. This means that the telescope does not need to waste time waiting for the dome to move when slewing to a new position, and can instead quickly switch from observing one portion of the sky to another with minimal dead time. Each GOTO site will host two independent mounts in separate domes, with a total of 16 unit telescopes giving an instantaneous field of view of approximately 80~square~degrees. Two sites are planned, one in either hemisphere, which will allow the entire sky to be observed.

\begin{figure}[p]
    \begin{center}
        \includegraphics[width=0.8\linewidth]{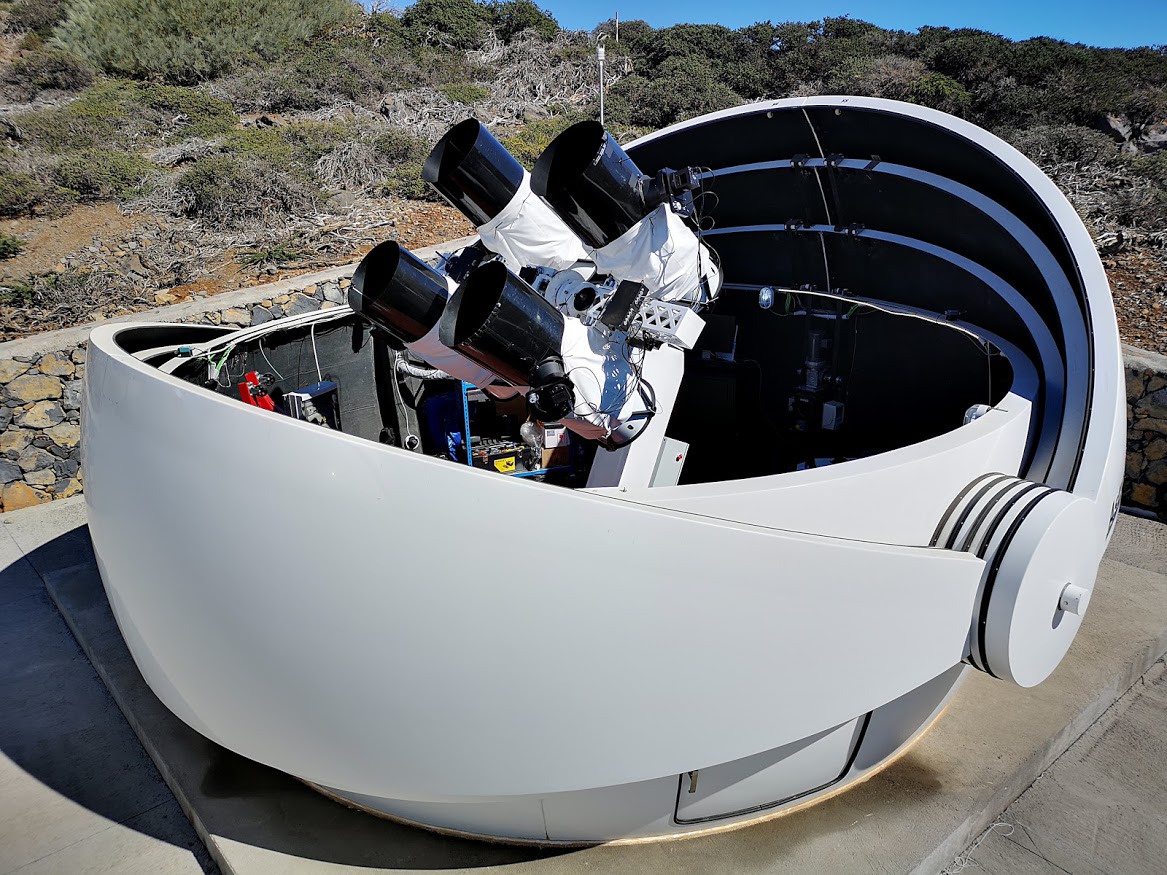}
    \end{center}
    \caption[example]{
        The GOTO-4 prototype on La Palma, pictured in 2019. The initial four unit telescopes are shown attached to the boom-arm of the prototype mount, all housed in an Astrohaven clamshell dome.
    }\label{fig:goto_photo_v1}
\end{figure}

\begin{figure}[p]
    \begin{center}
        \includegraphics[width=0.8\linewidth]{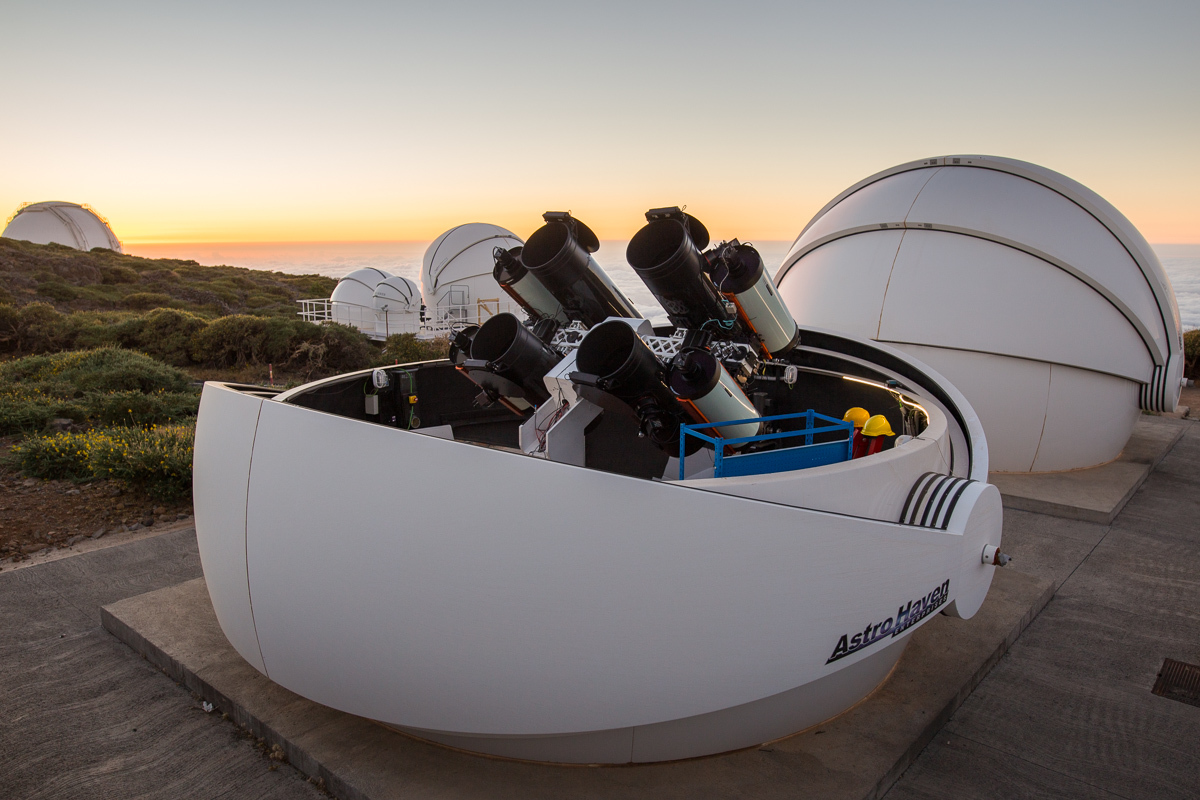}
    \end{center}
    \caption[example]{
        The GOTO configuration since August 2020. The prototype unit telescopes have been replaced with new UTs, and four Celestron RASAs have been added to the outside of the array. The second dome is also visible on the right.
    }\label{fig:goto_photo_v2}
\end{figure}

The first GOTO prototype was built by APM Professional Telescopes, and was deployed at the Observatorio del Roque de los Muchachos on La Palma in the Canary Islands in the summer of 2017 (shown in Figure~\ref{fig:goto_photo_v1}). The prototype was commissioned with four first-generation unit telescopes, and during the commissioning period several design improvements were identified for both the UTs and the mount. For the unit telescopes the major issues were the stability of the guidemounts connecting each OTA to the boom-arm, as well as the need to add baffles and covers to reduce scattered light (visible in Figure~\ref{fig:goto_photo_v1}). The mount motor mechanism was also upgraded, and additional switches and sensors were added to the dome to provide extra information when operating robotically. In 2020 the first second-generation unit telescopes built by ASA Astrosysteme were added to the mount, replacing the original four in the centre of the array (shown in Figure~\ref{fig:goto_photo_v2}). The major change between the two generations was the switch from open to closed tubes. Four additional Celestron Rowe-Ackermann Schmidt Astrographs (RASAs) were also added to the outside of the array, as temporary units while the original OTAs are refurbished (also visible in Figure~\ref{fig:goto_photo_v2}). A second dome has also been constructed on the La Palma site to host the second mount, which is due to be added in early 2021 (see Section~\ref{sec:future}).

\newpage

\clearpage

\subsection{Software}
\label{sec:software}

\begin{figure}[t]
    \begin{center}
        \includegraphics[width=0.8\linewidth]{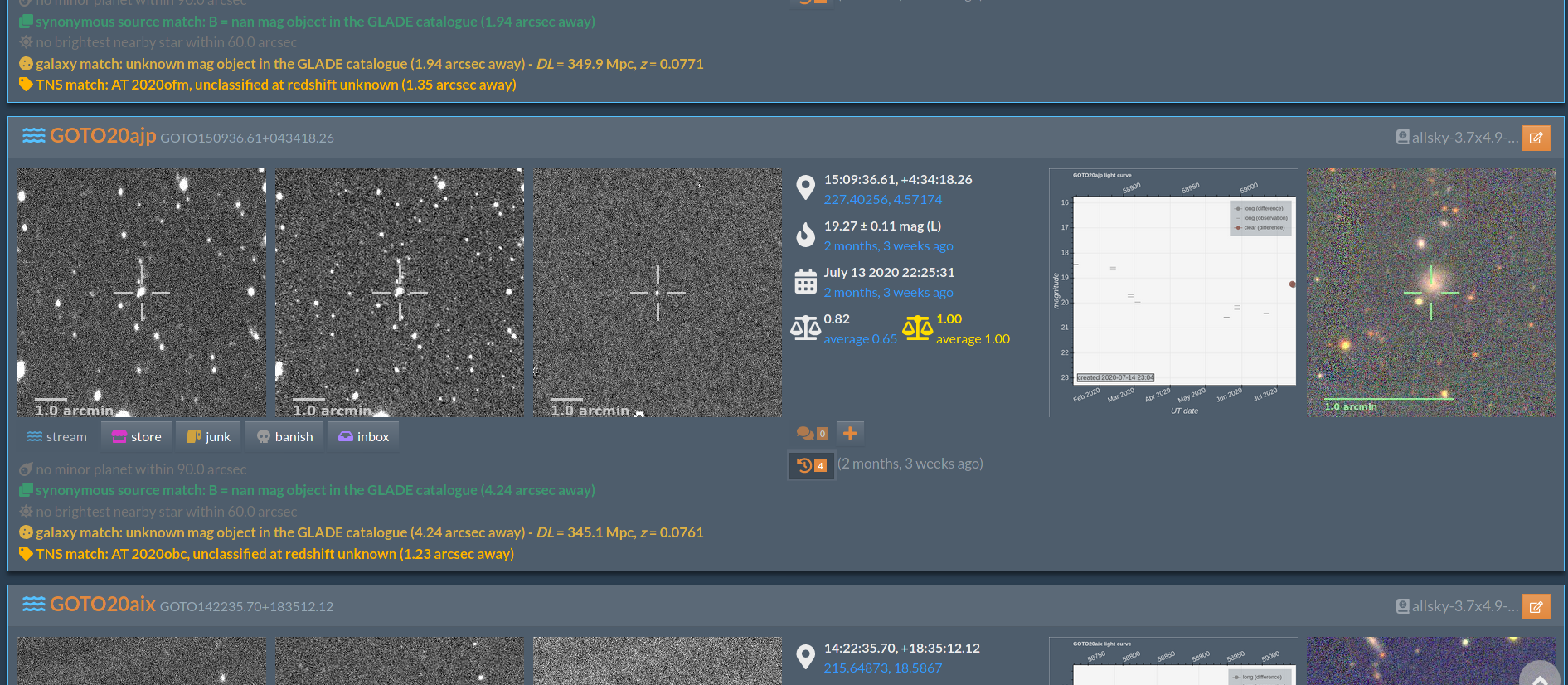}
    \end{center}
    \caption[example]{
        An entry for a candidate in the GOTO Marshall. From left to right the detection, reference and difference images are shown, as well as context information including a light curve,  matches to galaxy catalogues and TNS entries.
    }\label{fig:marshall}
\end{figure}

The GOTO hardware is operated using a custom robotic control system, the GOTO Telescope Control System (G\nobreakdash-TeCS)\cite{gtecs,thesis}. G\nobreakdash-TeCS is written in Python and is based around a series of hardware control daemons, which are monitored and sent commands by the pilot master control program. The pilot replicates the role of a human observer: carrying out tasks and observing throughout the night, while also monitoring the hardware and the site conditions. If a hardware error is detected then the pilot will run through a series of pre-defined recovery commands to try and fix it, and if the weather conditions turn poor then the dome will automatically close until they are good again.

Pointings are scheduled and communicated to the pilot by the G\nobreakdash-TeCS scheduler, which filters and sorts the targets added to a central observation database. The system alert listener monitors the NASA GCN steam \cite{GCN} for relevant astrophysical events (gravitational-wave alerts from the LIGO-Virgo Collaboration and gamma-ray burst events detected by \textit{Fermi} and \textit{Swift}) and adds targets to the database to cover the localisation region. The scheduler recalculates the highest priority pointing every 10 seconds, resulting in a system that is very quick to react to transient alerts as new targets are added to the database. During the prototype phase GOTO was able to start exposing observations of new gravitational-wave events within 30 seconds of the alert being received by the sentinel\cite{Gompertz2020}.

Once images are taken on La Palma they are transferred to the GOTO data centre at Warwick University in the UK through a dedicated fibre link. At the data centre the images are processed by the GOTOphoto data reduction pipeline. GOTOphoto first calibrates each individual image, using master bias, dark and flat frames which are updated on a monthly basis. Then initial source detection and astrometric calculations are carried out on each image, using SExtractor\cite{sextractor} and astrometry.net\cite{astrometrydotnet} respectively, before the detected sources are matched to the ATLAS-REFCAT2\cite{atlasrefcat} reference catalogue. A standard GOTO survey observation comprises a set of three 60~s exposures, taken using the \textit{L} filter (400--700~nm). After being processed individually the images from each set are aligned and combined and then reprocessed through the same methods outlined above, to allow for deeper detections as well as the detection of objects moving between the exposures. Each combined image is also subtracted from the appropriate master reference frame for that sky position using HOTPANTS \cite{hotpants}, and the resulting difference image is then used to detect transient sources.

Candidates are classified using a convolutional neural network, which assigns each source a score between "real" (1) and "bogus" (0)\cite{Mong2020,Killestein}. High-scoring sources are presented for human vetting through a web interface called the GOTO Marshall (see Figure~\ref{fig:marshall}). Collaboration members can check each candidate and flag them either as potential astrophysical sources or junk detections, based on historic light curves, image stamps and cross-matching with astrophysical catalogues. From the Marshall, authorised users can also schedule automated follow-up observations using other telescopes on La Palma, including \textit{pt5m}\cite{pt5m} and the Liverpool Telescope\cite{Liverpool}.

\newpage

\section{Deployment}
\label{sec:deployment}

\subsection{The GOTO-4 prototype}
\label{sec:prototype}

\begin{figure}[t]
    \begin{center}
        \includegraphics[width=0.95\linewidth]{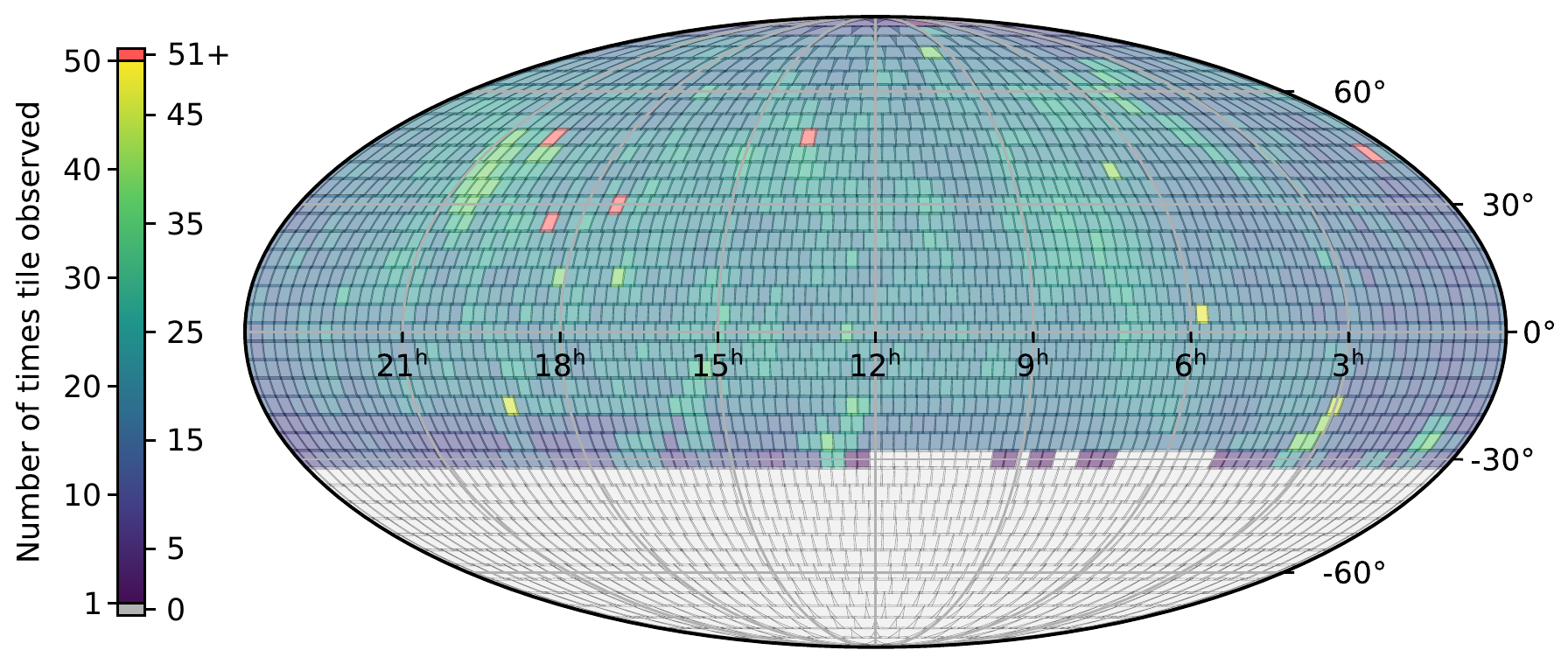}
    \end{center}
    \caption[example]{
        Tiles observed during the GOTO prototype period (February 2019 -- August 2020). Tiles in grey were never observed due to being below the visible horizon, while tiles in red were observed more frequently than average due to containing targets that were the subject of detailed campaigns (such as M31).
    }\label{fig:tiles}
\end{figure}

The GOTO prototype telescope was inaugurated on 3 July 2017 on La Palma. Over the next 18 months all aspects of the hardware and software were tested and commissioned, with the aim of achieving a functional and fully-automated system by the start of the third LIGO-Virgo gravitational-wave observing run (O3). The system left the commissioning stage and begun fully-automated survey operations on the 21 February 2019, just over a month before O3 began on 1 April. GOTO-4 followed up 52 of the 76 triggers published during O3 (including retractions), until the observing run was suspended on 23 March 2020 due to the COVID-19 pandemic. GOTO itself had to pause automated operations until the start of May due to local lockdowns on La Palma, but afterwards resumed regular observing until August 2020 when the unit telescopes were replaced (as described in Section~\ref{sec:hardware}). A complete map of grid tiles observed during the period from February 2019 to August 2020 is shown in Figure~\ref{fig:tiles}. Further details of the GOTO prototype phase will be presented in [\citeonline{GOTOprototype}].

\subsection{Future expansion}
\label{sec:future}

In 2019 the GOTO collaboration secured funding from the UK Science and Technology Facilities Council to construct a second full mount system with 8 unit telescopes on La Palma. Having two complete mounts at the same site will increase the instantaneous field of view to 80~square~degrees, halve the all-sky survey cadence from $\sim$5 to 2.5 days, and create new opportunities for follow-up scheduling (for instance observing the same tile with both mounts to increase the effective depth, or using two different filters simultaneously). The second system is currently under construction and is intended to be installed at the site in early 2021 (see Table~\ref{tab:timeline}).

Funding has also been secured for a southern GOTO node with at least one additional mount, to be located at Siding Spring Observatory in New South Wales, Australia. Adding a southern node will allow coverage of the entire visible sky, making it possible to cover more of the gravitational-wave localisation regions. As the Canary Islands and Australia are on opposite sides of the globe the second site will also enable close to 24-hour observations, ensuring that GOTO is able to react to alerts whenever they occur.

Ultimately all the telescopes across both sites will be operated as a single observatory. Although each telescope will have an independent robotic control system it is anticipated that they will receive commands from a single central scheduler, which will allow a rapid, coordinated response to any transient alerts. Likewise all image data will be sent back to the Warwick data centre and processed together, resulting in a near 24-hour stream of new images and candidate sources from the two antipodal sites.

\begin{table}[t]
\begin{center}
\begin{tabular}{llllll}
\textbf{Date} &
\textbf{Event} &
\multicolumn{3}{l}{\textbf{Hardware}} &
\textbf{Total field of view} \\
\hline
Jul 2017  & GOTO-4 prototype inaugurated      & GOTO-N: & 1 mount & 4 UTs   & 18~square~degrees \\
Aug 2020  & Prototype upgraded to 8 UTs       & GOTO-N: & 1 mount & 8 UTs   & 40~square~degrees \\
\textit{Early 2021} & Second mount installed  & GOTO-N: & 2 mounts & 16 UTs & 80~square~degrees \\
\textit{Mid 2021}   & GOTO-South inaugurated  & GOTO-N: & 2 mounts & 16 UTs & 120~square~degrees \\
                    &                         & GOTO-S: & 1 mount & 8 UTs  & \\
\textit{2022} & Second Australian mount added & GOTO-N: & 2 mounts & 16 UTs & 160~square~degrees \\
                    &                         & GOTO-S: & 2 mounts & 16 UTs &  \\
\end{tabular}
\end{center}
\caption{Timeline of stages in the GOTO project. Dates in italics are current projections. Hardware is divided between La Palma in the north (GOTO-N) and Siding Spring in the south (GOTO-S).}
\label{tab:timeline}
\end{table}

\newpage
\section{Conclusion}
\label{sec:conclusion}

We have presented the GOTO project, which aims to build a network of wide-field robotic telescopes across the globe dedicated to detecting electromagnetic counterparts to gravitational-wave events. Each GOTO mount holds an array of eight 40~cm unit telescopes, which combine to a total field of view of 40~square~degrees, and can observe to $\sim$20~mag in a set of three 60~s exposures.

The first GOTO prototype telescope has been operational on La Palma since 2017, and is in the process of being upgraded with new hardware. A second telescope is planned to be commissioned at the same site in early 2021, with two further mount systems planned for a southern node in Australia later in the year. The four telescopes will form a global network, able to survey the entire sky every 2-3 days, and will be ready to react to new gravitational-wave alerts once the next gravitational-wave observing run begins in 2022.

\section*{Acknowledgements}

The Gravitational-wave Optical Transient Observer (GOTO) project acknowledges the support of the Monash-Warwick Alliance; the University of Warwick; Monash University; the University of Sheffield; the University of Leicester; Armagh Observatory \& Planetarium; the National Astronomical Research Institute of Thailand (NARIT); the Instituto de Astrofísica de Canarias; the University of Turku; the University of Manchester; the University of Portsmouth and the UK Science and Technology Facilities Council (grant numbers ST/T007184/1 and ST/T003103/1).

\bibliography{report}
\bibliographystyle{spiebib_shortlist}


\end{document}